\documentclass
[prd,onecolumn,showpacs,floatfix,superscriptaddress,12pt]{revtex4}%
\usepackage{graphicx}
\usepackage{subfigure}
\usepackage{epsf}
\usepackage{bm}
\usepackage{amsmath}
\usepackage{amsfonts}
\usepackage{amssymb}
\usepackage{graphicx}
\usepackage{tabularx}
\usepackage{color}%
\providecommand{\U}[1]{\protect\rule{.1in}{.1in}}

\newcommand{\be}{\begin{equation}}
\newcommand{\ee}{\end{equation}}

\newcommand{\mincir}{\raise
-3.truept\hbox{\rlap{\hbox{$\sim$}}\raise4.truept\hbox{$<$}\ }}
\newcommand{\magcir}{\raise
-3.truept\hbox{\rlap{\hbox{$\sim$}}\raise4.truept\hbox{$>$}\ }}

\begin{document}
\title{Classical and Quantum solutions in Scalar field cosmology via the Eisenhart
lift and linearization}
\author{Andronikos Paliathanasis}
\email{anpaliat@phys.uoa.gr}
\affiliation{Institute of Systems Science, Durban University of Technology, PO Box 1334,
Durban 4000, South Africa}
\affiliation{Departamento de Matem\'{a}ticas, Universidad Cat\'{o}lica del Norte, Avda.
Angamos 0610, Casilla 1280 Antofagasta, Chile}

\begin{abstract}
This study introduces a novel approach for solving the cosmological field
equations within scalar field theory by employing the Eisenhart lift. The
field equations are reformulated as a system of geodesic equations for the
Eisenhart metric. In the case of an exponential potential, the Eisenhart
metric is shown to be conformally flat. By applying basic geometric
principles, a new set of dynamical variables is identified, allowing for the
linearization of the field equations and the derivation of classical
cosmological solutions. However, the quantization of the Eisenhart system
reveals a distinct set of solutions for the wavefunction, particularly in the
presence of symmetry breaking at the quantum level.

\end{abstract}
\keywords{Quantum Cosmology; Scalar Field; Eisenhart lift; Wheeler-DeWitt equation}
\pacs{98.80.-k, 95.35.+d, 95.36.+x}
\date{\today}
\maketitle

\section{Introduction}

Gravitational models with scalar fields play a crucial role in describing
cosmological observations \cite{ameb10}. Specifically, scalar fields provide
new degrees of freedom in the Einstein field equations, influencing the
dynamics to reproduce the observed cosmological history
\cite{dataacc1,dataacc2,data1,data2,Hinshaw:2012aka,Ade:2015xua}.
Additionally, scalar fields attributes the new degrees of freedom provided by
the modification of the Einstein-Hilbert Action with the introduction of new
geometric invariants \cite{soti}.

In the literature, numerous scalar field models have been proposed; for
example, refer to \cite{f1,f2,f3,f4,f5,f6,f7,f8,f9,f10,f11} and the references
therein. Quintessence represents the simplest among these scalar field models.
In this model, the scalar field is minimally coupled to gravity, featuring a
potential function to describe the scalar field mass. Quintessence provides a
straightforward mechanism for cosmic acceleration \cite{Ratra}. The universe
experiences acceleration when the potential function dominates over the
kinetic term. Due to its simplicity, the quintessence model has been applied
to describe various cosmological phases, including the early-time acceleration
phase, inflation \cite{guth,Aref1,de0}, and the late-time acceleration
associated with dark energy \cite{de1,de2,de3,de4}. Moreover, the quintessence
model has been utilized as a framework to unify the dark sector of the
universe, encompassing both dark energy and dark matter \cite{ur1,ur2,ur3}.

While the field equations of quintessence cosmology are second-order, they are
nonlinear. Unfortunately, nonlinear differential equations, in general, are
not conducive to analytic treatment. Typically, they can be addressed using
numerical methods; however, such methods may not always yield a comprehensive
understanding of the solution's behavior across the entire solution space. On
the other hand, the asymptotic behavior of the field equations can be
examined, yet even in this scenario, information about the solution's behavior
far from the asymptotic points remains elusive.

A systematic method that has been employed to solve nonlinear field equations
is symmetry analysis \cite{lie1,lie2,lie3}. In this approach, the demand for
field equations to be invariant under certain transformations leads to the
identification of conservation laws. These conservation laws have been
utilized to treat cosmological field equations as systems of algebraic
equations, this approach applied before in various cosmological models
\cite{ns1,ns2,ns3,ns4}

In this study, we focus on the application of the Eisenhart lift
\cite{el1,el2,el3} to analyze cosmological field equations, specifically for
the quintessence model. The field equations of the quintessence model allow
for a minisuperspace description, forming a two-dimensional Hamiltonian system
with the scale factor and the scalar field as dynamical variables under a
potential function. The Eisenhart approach involves increasing the dimension
of the Hamiltonian system, allowing us to express the equations of motion as a
set of geodesic equations. The resulting metric is referred to as the
Eisenhart metric. We study the geometric properties for the Eisenhart metric
and we show how the field equations can be linearized. In this case we are
able to solve the field equations through linearization. Moreover, except from
the classical solution we focus and on the quantization and we derive the
extended Wheeler-DeWitt equation \cite{wdw1,wdw2} of quantum cosmology
\cite{kim,pinto,manto,Stu2,sing1,qq1,qq2,qq3}. 

The Eisenhart lift for field theories has been explored in \cite{pap0}, and
the quantization process of the Eisenhart lift was discussed in \cite{pap0a}.
Recent investigations of the Eisenhart lift for the quintessence model are
presented in \cite{pap1}, covering both the Eisenhart lift and the
quantization process. In a broader context, the Eisenhart lift has been
applied to analyze cosmological field equations in \cite{pap2}. While an
alternative approach involving the Jacobi metric has been discussed in
\cite{pap3}, our focus in this work remains on the Eisenhart lift. The
structure of the paper is as follows.

In Section \ref{sec2}, we present the gravitational model under consideration,
which is that of a Friedmann--Lema\^{\i}tre--Robertson--Walker (FLRW) geometry
with a minimally coupled scalar field. The field equations admit a
minisuperspace description, and we derive the Wheeler-DeWitt (WDW) equation
for quantum cosmology. In Section \ref{sec3}, we discuss the basic
mathematical elements for the Eisenhart lift. Section \ref{sec4} reviews some
previous results on the symmetries and conservation laws for the geodesic
equations. Additionally, for the exponential potential, the Eisenhart metric
for the scalar field model is three-dimensional, and it possesses the property
that the Cotton-York tensor is always zero. Thus, the field equations, which
are invariant under conformal transformations, can be expressed in the form of
three free particles. Indeed, the requirement for the Eisenhart metric to be
conformally flat is equivalent to the requirement for the field equations to
be linearized. The classical and quantum solutions of the linearized system
are presented in Section \ref{sec5}. Finally, in Section \ref{sec6}, we
summarize our results and draw conclusions.

\section{Scalar field cosmology}

\label{sec2}

In Einstein's General Relativity, the contribution of the matter source to the
curvature of the physical space with metric $g_{\mu\nu}$ is given by the field
equations \cite{ameb10}
\begin{equation}
R_{\mu\nu}-\frac{1}{2}g_{\mu\nu}R=k\ T_{\mu\nu},\label{EE}%
\end{equation}
in which the left-hand side of expression (\ref{EE}) corresponds to the
Einstein tensor described by the Ricci tensor $R_{\mu\nu}$, which includes the
Ricci tensor and Ricci scalar $R$. The right-hand side of expression
(\ref{EE}) gives the contribution of the matter source in the field equations,
where $T_{\mu\nu}$ is the effective energy-momentum tensor, and its equation
of motion is $T_{~~;\nu}^{\mu\nu}=0$. The parameter $k$ is the Einstein
constant, defined as $k=8\pi G$. Without loss of generality, in the following,
we consider that $k=1$.

In very large scales, the universe is described by the homogeneous and
isotropic FLRW spacetime with the line element%
\begin{equation}
ds^{2}=-dt^{2}+a^{2}(t)\frac{1}{(1+\frac{\mathit{K}}{4}\mathbf{x}^{2})^{2}%
}(dx^{2}+dy^{2}+dz^{2}), \label{SF.1}%
\end{equation}
where $a(t)$ is the scale factor of the universe and $\mathit{K}=0,\pm1$ is
the spatial curvature parameter.

For the line element (\ref{SF.1}), Einstein's field equations (\ref{EE}) are
reduced from a system of nonlinear second-order partial differential equations
to the following system of nonlinear second-order ordinary differential
equations \cite{ameb10}%
\begin{equation}
H^{2}+\frac{\mathit{K}}{a^{2}}=\frac{1}{3}\rho,\label{fr.01}%
\end{equation}%
\begin{equation}
2\dot{H}+3H^{2}+\frac{\mathit{K}}{a^{2}}=-p,\label{fr.02}%
\end{equation}
where $\rho$ and $p$ are the energy density and pressure components for the
fluid, that is
\begin{equation}
T_{\mu\nu}=\left(  \rho+p\right)  u_{\mu}u_{\nu}+pg_{\mu\nu},
\end{equation}
the vector field $u_{\mu}$ is the comoving observer, $u^{\mu}u_{\mu}=-1$, and
$H$ is the Hubble function, $H=\frac{\dot{a}}{a}$, related to the expansion
rate $\theta=3H,$ $\theta=u_{;\mu}^{\mu}$.

The equation of motion for the matter source reads%
\begin{equation}
\dot{\rho}+3H\left(  \rho+p\right)  =0. \label{em1}%
\end{equation}

Assume now that the energy-momentum tensor $T_{\mu\nu}$ attributes the degrees
of freedom of a scalar field minimally coupled to gravity, which inherits the
symmetries of the background geometry. That is, \cite{Ratra}
\begin{equation}
T_{\mu\nu}=-\frac{2}{\sqrt{-g}}\frac{\delta(\sqrt{-g}L_{\phi})}{\delta
g^{\mu\nu}},
\end{equation}
where
\begin{equation}
L_{\phi}\left(  \phi,\phi_{;\nu}\right)  =-\frac{1}{2}g^{\mu\nu}\phi_{,\mu
}\phi_{,\nu}-V(\phi),
\end{equation}
is the scalar field Lagrangian, $\phi=\phi\left(  t\right)  $, $\phi_{,\nu
}=\dot{\phi}\delta_{\nu}^{0}$, and $V\left(  \phi\right)  $ is the scalar
field potential.

The energy density and pressure components are defined as \cite{Ratra}
\begin{equation}
\rho\equiv\frac{1}{2}\dot{\phi}^{2}+V(\phi)~,\label{fr.03}%
\end{equation}%
\begin{equation}
p=\frac{1}{2}\dot{\phi}^{2}-V(\phi)~,\label{fr.04}%
\end{equation}
in which \cite{Ratra}
\begin{equation}
w_{\phi}=\frac{\frac{1}{2}\dot{\phi}^{2}-V(\phi)}{\frac{1}{2}\dot{\phi}%
^{2}+V(\phi)},
\end{equation}
correspond to the equation of state parameter. In the limit where the scalar
field is negligible with respect to the potential energy, that is, $\frac
{\dot{\phi}^{2}}{2}\ll V(\phi)$, then the equation of state parameter reaches
the asymptotic limit $w_{\phi}\simeq-1~$\cite{Ratra}.

Finally, the equation of motion (\ref{em1}) for the matter source is written
in the form of the Klein-Gordon equation
\begin{equation}
\ddot{\phi}+3H\dot{\phi}+V_{,\phi}=0. \label{fr.05}%
\end{equation}

According to cosmological observations, the spatial curvature is zero; thus,
in the following, we assume $\mathit{K~}=0$.

\subsection{Minisuperspace description}

The cosmological field equations (\ref{fr.01}), (\ref{fr.02}), (\ref{fr.03}),
(\ref{fr.04}) and (\ref{fr.05}) have the property to admit a minisuperspace
description. Specifically, there exist the point-like Lagrangian function
\cite{ns1}
\begin{equation}
L\left(  a,\dot{a},\phi,\dot{\phi}\right)  =-3a\dot{a}^{2}+\frac{1}{2}%
a^{3}\dot{\phi}^{2}-a^{3}V\left(  \phi\right)  ,\label{lang.01}%
\end{equation}
where variation with respect to dynamical variables $a$ and $\phi$ gives the
second-order field equations. The constraint equation (\ref{fr.01}) can be
seen as the Hamiltonian of the Lagrangian function (\ref{lang.01}).

On the other hand, if we introduce the lapse function $N\left(  t\right)  $ in
the metric tensor (\ref{SF.1}), that is $dt\rightarrow N\left(  t\right)  dt$,
the point-like Lagrangian is singular, that is,%
\begin{equation}
\bar{L}\left(  N,a,\dot{a},\phi,\dot{\phi}\right)  =\frac{1}{N}\left(
-3a\dot{a}^{2}+\frac{1}{2}a^{3}\dot{\phi}^{2}\right)  -Na^{3}V\left(
\phi\right)  , \label{sinl.01}%
\end{equation}
in which now the constraint equation (\ref{fr.01}) follows from the variation
of (\ref{sinl.01}) with respect to the lapse function $N$, that is, the
constraint equation reads $\frac{\partial L}{\partial N}=0$.

From the point-like Lagrangian we define the two-dimensional minisuperspace
$\gamma_{ij}$ with diagonal line element%
\begin{equation}
ds_{m}^{2}=-6ada^{2}+a^{3}d\phi^{2}. \label{mn.01}%
\end{equation}

We introduce the Legendre transformation, $p_{a}=\frac{\partial L}%
{\partial\dot{a}},~p_{\phi}=\frac{\partial L}{\partial\dot{\phi}},$ where
$p_{a}$ and $p_{\phi}$ are the momentum, thenthe field equations read
\cite{ns1}
\begin{equation}
-\frac{1}{12a}p_{a}^{2}+\frac{1}{2a^{3}}p_{\phi}^{2}+a^{3}V\left(
\phi\right)  =0.\label{lang.02}%
\end{equation}%
\begin{equation}
\dot{a}=-\frac{1}{6a}\dot{p}_{a}~,~\dot{\phi}=\frac{1}{a^{3}}p_{\phi}%
~,~\dot{p}_{\phi}=-a^{3}V_{,\phi},
\end{equation}%
\begin{equation}
\dot{p}_{a}=-\frac{1}{12}\frac{p_{a}^{2}}{a^{2}}+\frac{3}{2}\frac{p_{\phi}%
^{2}}{a^{4}}-3a^{2}V\left(  \phi\right)  .
\end{equation}
or, in the presence of the lapse function
\begin{equation}
\frac{1}{N}\left(  -\frac{1}{12a}p_{a}^{2}+\frac{1}{2a^{3}}p_{\phi}%
^{2}\right)  +Na^{3}V\left(  \phi\right)  =0.\label{mn.01aa}%
\end{equation}%
\begin{equation}
\frac{1}{N}\dot{a}=-\frac{1}{6a}p_{a}~,~\frac{1}{N}\dot{\phi}=\frac{1}{a^{3}%
}p_{\phi}~,~\frac{1}{N}\dot{p}_{\phi}=-a^{3}V_{,\phi},
\end{equation}%
\begin{equation}
\frac{1}{N}\dot{p}_{a}=-\frac{1}{12}\frac{p_{a}^{2}}{a^{2}}+\frac{3}{2}%
\frac{p_{\phi}^{2}}{a^{4}}-3a^{2}V\left(  \phi\right)  .\label{mn.01a}%
\end{equation}

\subsection{Quantization}

In the context of the $3+1$ decomposition notation, the WDW equation is
derived from the Hamiltonian constraint of General Relativity and can be
expressed as follows \cite{wdw1,wdw2}%

\begin{equation}
\mathcal{H}\Psi=\left[  -4\kappa^{2}\mathcal{G}_{ijkl}\frac{\delta^{2}}{\delta
h_{ij}\delta h_{kl}}+\frac{\sqrt{h}}{4\kappa^{2}}\left(  -\mathcal{R}%
+2\Lambda+4\kappa^{2}T^{00}\right)  \right]  \Psi=0, \label{WDW1}%
\end{equation}
where%
\begin{equation}
\mathcal{G}_{ijkl}=\frac{1}{2\sqrt{h}}\left(  h_{\mu\kappa}h_{\nu\lambda
}+h_{\mu\lambda}h_{\nu\kappa}-h_{\mu\nu}h_{\kappa\lambda}\right)  ,
\label{WDW2}%
\end{equation}
is the the metric of superspace, that is, the space of all 3-geometries, with
metric $h_{ij}$ and Ricci scalar $R$, and the matter configuration.

In the case of a scalar field we find%

\begin{equation}
T^{00}=-\frac{1}{2h}\frac{\delta^{2}}{\delta\phi^{2}}+\frac{1}{2}h^{\mu\nu
}\phi_{,\mu}\Phi_{,\nu}+V(\Phi). \label{WDW3}%
\end{equation}
We note that equation (\ref{WDW1}) does not represent a single differential
equation but rather a family of equations on the -dimensional hypersurface.
However, in the presence of a minisuperspace, the infinite degrees of freedom
of the superspace are reduced to a finite number, leading to the
simplification of the WDW equation into a single equation.

The Hamiltonian constraint (\ref{mn.01aa}) reads \ \cite{wdw1,wdw2}%
\begin{equation}
\mathcal{H}\equiv\frac{1}{N^{2}}\left(  \gamma_{ij}p^{i}p^{j}\right)
+2\tilde{V}=0\label{mn.02}%
\end{equation}
and the WDW equation follows from the quantization of (\ref{mn.02}), that is,
\[
\overset{\symbol{94}}{\mathcal{H}}\Psi=0.
\]

Yet, since the field equations remain invariant regardless of the chosen lapse
function, which functions as a conformal factor, we ensure the invariance of
the WDW equation under conformal transformations. Instead of employing the
typical Laplace operator, we utilize the conformal Laplace operator%

\begin{equation}
\hat{L}\equiv-\Delta-\frac{n-2}{4\left(  n-1\right)  }R_{\gamma},
\end{equation}
in which $\Delta=~\frac{1}{\sqrt{\left\vert \gamma\right\vert }}\frac
{\partial}{\partial x^{i}}\left(  \sqrt{\left\vert \gamma\right\vert }%
\frac{\partial}{\partial x^{j}}\right)  $ is the Laplacian operator,
$n=\dim\gamma_{ij}$, and $R_{_{\gamma}}$ is the Ricci scalar for the
minisuperspace. In the case of two-dimensional minisuperspace it follows
$\hat{L}=\Delta.$

Thus, for the minisupespace (\ref{mn.01}), we write the WDW equation
\cite{qq1}
\begin{equation}
\Delta\Psi-2a^{3}V\left(  \phi\right)  \Psi=0,
\end{equation}
in which operator $\Delta$ is defined as%
\begin{equation}
\Delta\equiv-\frac{1}{6a}\left(  \frac{\partial^{2}}{\partial a^{2}}%
+\frac{\partial}{\partial a}\right)  +\frac{1}{a^{3}}\frac{\partial^{2}%
}{\partial\phi^{2}}.\label{lang.06}%
\end{equation}

\section{The Eisenhart lift}

\label{sec3}

The Jacobi metric and the Eisenhart lift delineate distinct methods for
geometrically representing dynamical systems. Notably, autonomous dynamical
systems can be formulated as a set of geodesic equations. In this
investigation, we focus on the Eisenhart lift, a technique that involves
augmenting the dimensionality of the dynamical system. Specifically, this
geometrization process entails introducing additional dimensions through the
inclusion of new dependent variables. A novel kinetic metric is introduced,
characterized by at least one isometry associated with a Noetherian
conservation. When this isometry is applied, the geodesic equations are
reduced back to the original dynamical system.

To demonstrate this, we consider the Hamiltonian
\[
H_{n}\equiv\frac{1}{2}\gamma^{ij}\left(  q\right)  p_{i}p_{j}+V\left(
q\right)  -h_{n}=0,
\]
with equations of motion%
\begin{equation}
\dot{q}^{i}=\gamma^{ij}p_{j}~,~\dot{p}_{i}=-\frac{1}{2}\gamma_{~~,i}^{jk}%
p_{j}p_{k}-V_{,i}.
\end{equation}

\subsection{Riemannian lift}

We introduce the new Hamiltonian \cite{eisn}%
\begin{equation}
H_{n+1}=\frac{1}{2}\gamma^{ij}\left(  q\right)  p_{i}p_{j}+\frac{1}{2}\alpha
V\left(  q\right)  p_{z}^{2}-h_{n+1}=0
\end{equation}
with equations of motion%
\begin{equation}
\dot{q}^{i}=\gamma^{ij}p_{j}~,~\dot{z}=\alpha V\left(  q\right)  p_{z},
\end{equation}
and%
\begin{equation}
\dot{p}_{i}=-\frac{1}{2}\gamma_{~~,i}^{jk}p_{j}p_{k}-\frac{1}{2}aV_{,i}%
p_{z}^{2}~,~\dot{p}_{z}=0.
\end{equation}
From the latter we defin the conservation law $p_{z}=p_{z0}$. The dynamical
system described by the Hamiltonian equation $H_{n}$ is recovered from the
Hamiltonian $H_{n+1}$, when $\frac{1}{2}ap_{z}^{2}=1$ and $h_{n+1}=h_{n}.$

The new $n+1$ metric with line element
\begin{equation}
ds_{EL\left(  n+1\right)  }^{2}=\gamma_{ij}dq^{i}dq^{j}+\frac{1}{\alpha
V\left(  q\right)  }dz^{2},
\end{equation}
is the Eisenhart metric. This lift is also known as Riemannian lift.

\subsection{Lorentzian lift}

An alternative apporach for the constuction of the Eisenhart metric is given
by the Lorentzian lift, also known as Eisenhart-Duval lift \cite{pap2}. In
this approach we consider the new Hamiltonian system \cite{eisn}%
\begin{equation}
H_{n+2}=\frac{1}{2}\gamma^{ij}\left(  q\right)  p_{i}p_{j}+V\left(  q\right)
p_{u}^{2}+p_{u}p_{v}-h_{n+2},
\end{equation}
where the Eisenhart metric is
\[
ds_{EL\left(  n+2\right)  }^{2}=\gamma_{ij}dq^{i}dq^{j}+2dudv-2Vdu^{2}\text{.}%
\]

The corresponding equations of motion are
\begin{equation}
\dot{q}^{i}=\gamma^{ij}p_{j}~,~\dot{u}=2V\left(  q\right)  p_{u}+p_{v}%
~,~\dot{v}=p_{v},
\end{equation}
and%
\begin{equation}
\dot{p}_{i}=-\frac{1}{2}\gamma_{~~,i}^{jk}p_{j}p_{k}-V_{,i}p_{u}^{2}~,~\dot
{p}_{u}=0,~\dot{p}_{v}=0\text{.}%
\end{equation}

Therefore, the dynamical system described by the Hamiltonian function $H_{n}$
is recovered from the latte equtions of motion for $p_{u}^{2}=1$ and
$p_{u}p_{v}-h_{n+2}=h_{n}$.

\section{Symmetries and conservation laws}

\label{sec4}

We proceed our discussion by presenting some significant findings regarding
the symmetries of the geodesic equations and the conformal Laplace operator.

\subsection{Noether symmetries of the geodesic Lagrangian}

Consider the geodesic Lagrangian
\begin{equation}
L\left(  q^{k},\dot{q}^{k}\right)  =\frac{1}{2}\gamma_{ij}\left(
q^{k}\right)  \dot{q}^{i}\dot{q}^{j}~,~\dot{q}^{i}=\frac{dq^{i}}%
{dt}\label{gl.01}%
\end{equation}
with \textquotedblleft energy\textquotedblright\ $\frac{1}{2}\gamma
_{ij}\left(  q^{k}\right)  \dot{q}^{i}\dot{q}^{j}=h$. Then under the Action of
the infinitesimal transformation%
\begin{align}
\bar{t} &  =t+\varepsilon\xi\left(  t,q^{k}\right)  ~,\\
\bar{q}^{i} &  =q^{i}+\varepsilon\eta^{i}\left(  t,q^{k}\right)  ~,
\end{align}
the variation $\delta S$ of the Action $S=\int L\left(  q^{k},\dot{q}%
^{k}\right)  dt$, remain invariant, if and only if, there exist a function
$f\left(  t,q^{k}\right)  $ such that \cite{noe}
\begin{equation}
X^{\left[  1\right]  }L+L\dot{\xi}=\dot{f},\label{gl.02}%
\end{equation}
where $X^{\left[  1\right]  }$ is the first extension for the generator
$X=\xi\left(  t,q^{k}\right)  \partial_{t}+\eta^{i}\left(  t,q^{k}\right)
\partial_{i}$ of the infinitesimal transformation~defined as%
\begin{equation}
X^{\left[  1\right]  }=X+\left(  \dot{\eta}^{i}-\dot{\xi}\dot{q}^{i}\right)
\partial_{\dot{q}^{i}}.
\end{equation}

The condition (\ref{gl.02}) for the variational symmetry is recognized as
Noether's first theorem, and the generator associated with the transformation
$X$ is identified as the Noether symmetry.

For a given Noether symmetry $X$, Noether's second theorem states that the
function%
\begin{equation}
\Phi\left(  X\right)  =\xi h-\frac{\partial L}{\partial\dot{q}^{i}}\eta^{i}+f.
\end{equation}
is a conservation law for the equations of motion.

The Noether symmetries for the geodesic Lagrangian (\ref{gl.01}) have been
widely studied in the literature. Specifically, it has been found that the
Noether symmetries for the geodesice Lagrangian (\ref{gl.01}) are constructed
by the elements of the Homothetic algebra for the metric tensor $\gamma
_{ij}\left(  q^{k}\right)  $, that is, by the Killing symmetries and the
Homothetic symmetry, for the exact mathematical relation we refer the reader
to \cite{noegrg}.

However, in the special case where the system is constraint, that is, $h=0$,
the geodesic equations are invariant under conformal transformations
\cite{palgrg}. As a result, the Noether symmetries are constructed by the
elements of the Conformal algebra of the metric tensor $\gamma_{ij}\left(
q^{k}\right)  $, that is, by the Killing symmetries, the Homothetic symmetry
and the Conformal symmetries. For this consideration, if $\eta^{i}\left(
q^{k}\right)  $ is a Conformal symmetry for the metric tensor $\gamma
_{ij}\left(  q^{k}\right)  $, then the quantity $\Phi\left(  \eta^{i}\right)
=\frac{\partial L}{\partial\dot{q}^{i}}\eta^{i}$ is a conservation law, for
more details see \cite{ndim} and references therein.

\subsection{Symmetries for the conformal Laplace operator}

Consider the Yamabe equation%
\begin{equation}
\left(  \Delta+\frac{n-2}{4\left(  n-1\right)  }R\right)  \Psi\left(
q^{i}\right)  =0, \label{yam.01}%
\end{equation}
in which $\Delta$ is the Laplacian for the metric tensor $\gamma_{ij}\left(
q^{k}\right)  $, with $n=\dim\gamma_{ij}$.

Let $Y=\eta^{i}\left(  q^{k},\Psi\right)  \partial_{i}+\zeta\left(  q^{k}%
,\Psi\right)  \partial_{\Psi}$ be the generator of the infinitesimal
transformation%
\begin{align}
\bar{q}^{i}  &  =q^{i}+\varepsilon\eta^{i}\left(  q^{k},\Psi\right)  ~,\\
\bar{\Psi}  &  =\Psi+\varepsilon\zeta\left(  q^{k},\Psi\right)  ~.
\end{align}
Then, equation (\ref{yam.01}) remain invariant if and only if \cite{yamabe}
\[
Y=\eta^{i}\left(  q^{k}\right)  \partial_{i}+\left(  \frac{2-n}{2}\psi\left(
q^{k}\right)  \Psi+a_{0}\Psi+b\left(  q^{k}\right)  \right)  \partial_{\Psi}%
\]
where $\eta^{i}\left(  q^{k}\right)  $ is a Conformal symmetry for the metric
tensor $\gamma_{ij}\left(  q^{k}\right)  $ with conformal factor $\psi\left(
q^{k}\right)  $, $a_{0}$ is a constant and $b\left(  q^{k}\right)  $ is an
arbitrary solution of equation (\ref{yam.01}). The latter two symmetry vectors
indicate that the Yamabe equation is linear. There exist a unique connection
between the symmetries of the Yamabe equation and the Noetherian conservation
laws for the null geodesic Lagrangian.

The maximum number of admitted conformal symmetries by the metric $\gamma
_{ij}\left(  q^{k}\right)  ~$are $\frac{1}{2}\left(  n+1\right)  \left(
n+2\right)  $. In that case the Yamabe equation will be characterized as
maximally symmetric. When the latter is true, $\gamma_{ij}\left(
q^{k}\right)  $ is conformally flat, and the Yamabe equation can be written as
the wave equation of the flat space, while the classical geodesic equations
can be linearized.

The criterio in order the tensor $\gamma_{ij}\left(  q^{k}\right)  $ depends
on the dimension of the space. Indeed, all two-dimensional spaces are
conformally flat \cite{ky}. A three-dimensional space is conformally flat when
the Cotton-York tensor \cite{ky}%
\begin{equation}
C_{\mu\nu\kappa}=R_{\mu\nu;\kappa}-R_{\kappa\nu;\mu}+\frac{1}{4}\left(
R_{;\nu}g_{\mu\kappa}-R_{;\kappa}g_{\mu\nu}\right)  ,~n=3
\end{equation}
is zero. Finally, for higher-dimensional spaces, the necessary condition in
order the metric tensor to be conformally flat is that the Weyl tensor be
zero, that is, \cite{ky}%
\begin{equation}
C_{\mu\nu\kappa\lambda}=R_{\mu\nu\kappa\lambda}-\frac{2}{n-2}\left(
g_{\mu\lbrack\kappa}R_{\lambda]\nu}-g_{\nu\lbrack\kappa}R_{\lambda]\mu
}\right)  +\frac{2}{\left(  n-1\right)  \left(  n-2\right)  }Rg_{\mu
\lbrack\kappa}g_{\lambda]\nu}.
\end{equation}

\section{The exponential potential}

\label{sec5}

We apply the Riemannian lift in order to write the gravitational field
equations of scalar field cosmology (\ref{fr.01}), (\ref{fr.02}),
(\ref{fr.03}), (\ref{fr.04}) and (\ref{fr.05}) as a set of geodesic equations.
We introduce the Hamiltonian function%
\begin{equation}
H_{RL}\equiv-\frac{1}{12a}p_{a}^{2}+\frac{1}{2a^{3}}p_{\phi}^{2}+\frac{1}%
{2}a^{3}V\left(  \phi\right)  p_{z}^{2}=0. \label{hm.01}%
\end{equation}
and Lagrangian function%
\begin{equation}
L_{RL}\left(  a,\dot{a},\phi,\dot{\phi},z,\dot{z}\right)  =-3a\dot{a}%
^{2}+\frac{1}{2}a^{3}\dot{\phi}^{2}+\frac{1}{2a^{3}V\left(  \phi\right)  }%
\dot{z}^{2}. \label{hm.02}%
\end{equation}

The three-dimensional minisuperspace has the following line element%
\begin{equation}
ds_{RL}^{2}=-6ada^{2}+a^{3}d\phi^{2}+\frac{1}{a^{3}V\left(  \phi\right)
}dz^{2}=0.
\end{equation}

The requirement the latter spacetime to be conformally flat gives the
constraint equations%
\begin{align}
V_{,\phi\phi}V-V_{,\phi}^{2}  &  =0,\\
V^{2}V_{,\phi\phi\phi}-4VV_{,\phi}V_{,\phi\phi}+3V_{,\phi}^{3}  &  =0.
\end{align}
Hence, it follows the unique potential%
\begin{equation}
V\left(  \phi\right)  =V_{0}e^{-\lambda\phi}.
\end{equation}
We remark that that for the exponential scalar field potential, the geodesic
equations described by the Hamiltonian (\ref{hm.01}) can be linearized.

\subsection{The classical solution through linearization}

We introduce the change of variables%
\begin{equation}
a=x^{\frac{1}{6-\sqrt{6}\lambda}}y^{\frac{1}{6+\sqrt{6}\lambda}}~,~\phi
=\frac{1}{\lambda^{2}-6}\left(  \left(  \sqrt{6}-\lambda\right)  \ln y-\left(
\sqrt{6}+\lambda\right)  \ln x\right)  .
\end{equation}

In the new coordinates the\ geodesic Lagrangian (\ref{hm.02}) becomes%
\begin{equation}
L_{RL}\left(  x,\dot{x},y,\dot{y},z,\dot{z}\right)  =x^{1+\frac{3}{\sqrt
{6}\lambda-6}}y^{1-\frac{3}{\sqrt{6}\lambda+6}}\left(  \frac{2}{\lambda^{2}%
-6}\dot{x}\dot{y}+\frac{1}{2V_{0}}\dot{z}^{2}\right)  .
\end{equation}

We introduce the lapse function $N\left(  t\right)  =x^{1+\frac{3}{\sqrt
{6}\lambda-6}}y^{1-\frac{3}{\sqrt{6}\lambda+6}}$, that is$~N\left(  t\right)
=a^{3}e^{-\lambda\phi}$. Hence, the conformal Lagrangian reads%
\begin{equation}
\bar{L}_{RL}\left(  x,\dot{x},y,\dot{y},z,\dot{z}\right)  =\left(  \frac
{2}{\lambda^{2}-6}\dot{x}\dot{y}+\frac{1}{2V_{0}}\dot{z}^{2}\right)  .
\label{hm.03a}%
\end{equation}

The cosmological field equations are%
\begin{equation}
\ddot{x}=0~,~\ddot{y}=0~,~\ddot{z}=0,
\end{equation}
with constraint equation%
\begin{equation}
\left(  \frac{2}{\lambda^{2}-6}\acute{x}\dot{y}+\frac{1}{2V_{0}}\dot{z}%
^{2}\right)  =0~,~\frac{1}{2}~\left(  \frac{1}{V_{0}}\dot{z}\right)
^{2}=1\text{.}%
\end{equation}

The analytic solution for quintessence model with exponential potential
derived before in \cite{Russo}. However, in this study we solve the field
equations through the linearization. 

On the other hand, for $\lambda=\sqrt{6}$, the transformation which linearizes
the field equations is%
\begin{equation}
a=x^{\frac{1}{12}}e^{\frac{y}{2}}~,~\phi=\sqrt{6}\left(  \frac{y}{2}-\frac
{1}{12}\ln x\right)  ,
\end{equation}
where the geodesic Lagrangian becomes%
\begin{equation}
\bar{L}_{RL}\left(  x,\dot{x},y,\dot{y},z,\dot{z}\right)  =-\frac{1}{2}%
\acute{x}\dot{y}+\frac{1}{2V_{0}}\dot{z}^{2}.\label{hm.03}%
\end{equation}

On the other hand, for $\lambda=-\sqrt{6}$ the corresponding transformation
is
\begin{equation}
a=6^{\frac{1}{6}}e^{\frac{x}{12}}y^{\frac{1}{12}}~,~\phi=\frac{\sqrt{6}}%
{12}\left(  \ln y-x\right)  ,
\end{equation}
where the resulting Lagrangian is again (\ref{hm.03}).

\subsection{New solutions for the WDW equation}

From the Lagrangian functions (\ref{hm.03a}) and (\ref{hm.03}) we derive the
WDW equation%
\begin{equation}
\Psi_{,xy}-\frac{1}{2}\bar{V}_{0}\Psi_{,zz}=0, \label{hm.04}%
\end{equation}
in which $\bar{V}_{0}=\frac{V_{0}}{2}$ for $\lambda^{2}=6$ or $\bar{V}%
_{0}=\frac{2}{\left(  \lambda^{2}-6\right)  }V_{0}$ for $\lambda^{2}\neq6$.

Equation (\ref{hm.04}) is the wave equation for the three-dimensional flat
space and it is maximally symmetric. Except from the trivial symmetries
$Y^{0}=\Psi\partial_{\Psi}$ and $Y^{\inf}=b\left(  x,y,z\right)
\partial_{\Psi}$, equation (\ref{hm.04}) admits ten symmetry vectors generated
by the ten Conformal symmetries of the flat space.

The ten symmetry vectors are expressed as follows%
\[
Y^{1}=\partial_{x}~,~Y^{2}=\partial_{y}~,~Y^{3}=\partial_{z}~,~
\]%
\[
Y^{4}=x\partial_{x}-y\partial_{y}~,~Y^{5}=z\partial_{x}+\bar{V}_{0}%
y\partial_{z}~,~Y^{6}=z\partial_{y}+\bar{V}_{0}x\partial_{z}~,
\]%
\[
Y^{7}=x\partial_{x}+y\partial_{y}+z\partial_{z}~,
\]%
\[
Y^{8}=2\left(  x^{2}\partial_{x}+z^{2}\partial_{y}+2xz\partial_{z}\right)
-\bar{V}_{0}x\Psi\partial_{\Psi}~,
\]%
\[
Y^{9}=2\left(  y^{2}\partial_{y}+z^{2}\partial_{y}+2yz\partial_{z}\right)
-\bar{V}_{0}y\Psi\partial_{\Psi}~,
\]%
\[
Y^{10}=2z\left(  x\partial_{x}+y\partial_{y}\right)  +\left(  2xy\bar{V}%
_{0}+z^{2}\right)  -z\Psi\partial_{\Psi}~.
\]

The symmetry vectors keep invariant the differential equation and the solution
$\Psi$. Thus, if $Y$ is a symmetry vector it follows that $Y\Psi=0$.

From the vector field $Y^{3}-\alpha_{0}Y^{0}~$ $\left(  \partial_{z}%
-\alpha_{0}\Psi\partial_{\Psi}\right)  $, we define the operator $\partial
_{z}\Psi-\alpha_{0}\Psi=0$, which gives that $\Psi\left(  x,y,z\right)
=\Psi_{0}\left(  x,y\right)  e^{\alpha_{0}z}$. By replacing in (\ref{hm.04})
we end with the two-dimensional equation%
\begin{equation}
\Psi_{0,xy}-\frac{\alpha_{0}^{2}}{2}\bar{V}_{0}\Psi_{0}=0\label{hm.05}%
\end{equation}
which is nothing else that the WDW\ equation for the scalar field model with
exponential potential. The symmetry vectores of the latter equation are the
fields $Y^{1},~Y^{2}\,$and$~Y^{4}$. The application of these vector fields
leads to the the solutions for the WDW equation derived in \cite{anmpla}.
Nevertheless, the application of other symmetry vector fields lead to new
solutions for the WDW equation.

For instance from the vector fields $Y^{1}-\alpha_{0}Y^{0}$ and $Y^{6}%
-\beta_{0}Y^{0}$ we derive the closed-form solution%
\begin{equation}
\Psi\left(  x,y,z\right)  =\frac{\Psi_{0}}{\sqrt{y}}\exp\left(  \alpha
_{0}xy-\frac{\left(  \beta_{0}-\alpha_{0}z\right)  ^{2}}{2\alpha_{0}\bar
{V}_{0}}\right)  .
\end{equation}

On the other hand, from the symmetries $Y^{4}-\alpha_{0}Y^{0}$ and
$Y^{7}-\beta_{0}Y^{0}$ we end with the solution%
\begin{equation}
\Psi\left(  x,y,z\right)  =x^{\frac{\alpha_{0}-\beta_{0}}{2}}y^{\frac
{\alpha_{0}+\beta_{0}}{2}}\Psi_{1}\left(  \sigma\right)  ~,~\sigma=\frac
{z}{\sqrt{xy}},
\end{equation}
in which%
\begin{equation}
\left(  2\bar{V}_{0}-\sigma^{2}\right)  \Psi_{1,\sigma\sigma}-\sigma\left(
1+2\beta_{0}\right)  \Psi_{1,\sigma}+\left(  \alpha_{0}^{2}-\beta_{0}%
^{2}\right)  \Psi_{1}=0.
\end{equation}

In a similar way we can construct various solutions for the WDW equation
(\ref{hm.04}). For other solutions of the wave equation we refer the reader to
\cite{barb}.

\section{Conclusions}

\label{sec6}

In the classical limit, the existence of additional conservation laws for the
field equations in the Eisenhart metric helps us linearize the field
equations, allowing us to express analytical solutions in a simplified form.
However, the situation is different for the WDW equation.

In the Eisenhart metric, additional symmetry vectors exist, leading to the
derivation of new solutions for the universe's wavefunction. Notably, these
new solutions are derived without applying any initial or boundary conditions
for the partial differential equations. The necessity of the condition
$\left(  \partial_{z}-\alpha_{0}\Psi\partial_{\Psi}\right)  \Psi=0$ arises to
derive the WDW equation for the original scalar field model.

Nevertheless, the geodesic Lagrangian (\ref{hm.02}) and the scalar field
Lagrangian (\ref{sinl.01}) describe the same dynamical system, wherein the
potential terms are expressed as dynamical variables. Consequently, the
quantization process differs, as the Eisenhart metric introduces new dynamical
degrees of freedom. However, all the wavefunctions describe the same classical
system. The imposition of the condition $\left(  \partial_{z}-\alpha_{0}%
\Psi\partial_{\Psi}\right)  \Psi=0$ is solely based on the mathematical
requirement to restore the WDW equation for the original model. Thus, the
violation of this condition, which can be viewed as a symmetry breaking, leads
to new wavefunctions and potentially introduces new behaviors in the very
early stages of the universe. The possible new solution for the quantum
equation throught the Eisenhart lift has been discussed before \cite{pap0a},
and it was mentioned for the case of the quintessence scalar field in
\cite{pap1}. However, no new solutions derived.

In the pressence of an ideal gas with equation of state parameter
$p_{m}=\left(  \gamma-1\right)  \rho_{m}$ the cosmological Lagrangian becomes
\begin{equation}
L\left(  a,\dot{a},\phi,\dot{\phi}\right)  =-3a\dot{a}^{2}+\frac{1}{2}%
a^{3}\dot{\phi}^{2}-a^{3}V\left(  \phi\right)  -\rho_{m0}a^{-3\left(
\gamma-1\right)  }. \label{lang.010}%
\end{equation}

Hence, the Eisenhart lift leads to the three-dimensional metric with line
element%
\begin{equation}
ds_{EL}^{2}=-6ada^{2}+a^{3}d\phi^{2}+\frac{1}{a^{3}V\left(  \phi\right)
+\rho_{m0}a^{-3\left(  \gamma-1\right)  }}dz^{2}.
\end{equation}
The requirement the latter line element to be conformally flat gives the
scalar field potentials
\begin{equation}
V\left(  \phi\right)  =V_{0}\exp\left(  \pm\frac{\sqrt{6}}{2}\gamma
\phi\right)  .
\end{equation}
Thus, there exists a transformation, as before, where the field equations for
this specific scalar field potential can be linearized.

In this study, we have considered the linearization of the field equations as
an approach to construct solutions through the Eisenhart lift. However, this
is not the only approach that we can apply for the study of the dynamical system.

Although the Eisenhart lift is a mathematical method for solving nonlinear
dynamical systems, we demonstrate that it can offer new insights into the
physical properties of quantum solutions. In a future study, we plan to
investigate the physical properties of symmetry breaking for the WDW equation.

\textbf{Data Availability Statements:} Data sharing is not applicable to this
article as no datasets were generated or analyzed during the current study.

\begin{acknowledgments}
The author thanks the support of Vicerrector\'{\i}a de Investigaci\'{o}n y
Desarrollo Tecnol\'{o}gico (Vridt) at Universidad Cat\'{o}lica del Norte
through N\'{u}cleo de Investigaci\'{o}n Geometr\'{\i}a Diferencial y
Aplicaciones, Resoluci\'{o}n Vridt No - 096/2022 and 098/2022.
\end{acknowledgments}

\end{document}